\documentstyle[11pt,tables]{article}
\newcommand{\bfx}[1]{\mbox{\boldmath $#1$}}

\begin{document}
\begin{center}

{\large \bf $\bfx N\overline{\bfx N}$ Annihilation
in Large $\bfx N_{\bfx c}$ QCD
with $\bfx \rho$ and $\bfx \omega$ Mesons}

\vspace{1.5cm}

Yang Lu\footnote{email: yanglu@walet.physics.upenn.edu}
and R. D. Amado\footnote{email: amado@walet.physics.upenn.edu}\\
Department of Physics,\\
University of Pennsylvania, Philadelphia,
PA 19104, USA
\vspace{0.4cm}

(May 16, 1995)
\end{center}
\begin{abstract}
We use classical/large $N_c$ (number of colors) QCD to study
nucleon--antinucleon annihilation at rest. We include pion, rho
and omega fields in the classical dynamics, starting from the
nonlinear sigma model. We begin with a spherical blob of pionic
matter with zero baryon number and energy of twice the nucleon
mass. This evolves according to the classical dynamics to pion
and vector meson fields into the radiation zone. The radiation
fields are quantized using the method of coherent states modified
to include isospin and four momentum conservation. Empirical
information is extracted from that coherent state. We find
good agreement with data for single particle momentum and number
spectra, and with branching ratios  to many annihilation channels.
All this emerges with nearly no free parameters.
\end{abstract}

\begin{center}
PACS number(s): 13.75.Cs, 11.80.Gw
\end{center}

\newpage

\section{Introduction}
There is growing interest in using classical versions of
QCD  to model the dynamics in strongly interacting
processes leading to many pions.  The notion is that the
many pion state can be approximated as classical pion radiation.
Classical QCD is QCD in the limit of a large number of colors
\cite{'tHooft,Witten},
($N_C$),
and is naturally expressed
in terms of the classical pion field. The strong interaction dynamics
is much easier in the classical than in the quantum
guise, and the pion radiation can be quantized in the
radiation zone using the method of coherent states.
This general approach of classical dynamics leading to
pion radiation that can then be quantized
has been applied to the evolution and
decay of the disoriented chiral condensate \cite{DCC},
to pion production in heavy ion collisions \cite{B&D}, and
to nucleon antinucleon annihilation \cite{ACDLS,bs&rda}.

Though no one has actually
derived a classical pion field theory directly form QCD, it
is generally agreed that any such theory will feature baryons
as nonperturbative topological ``knots" of the pion
field and corresponding topologically stable baryon number.
Skyrme \cite{Skyrme} developed just such a theory long before the
advent of QCD. Some of its details may be special to Skyrme's
theory, but its general features should pertain to any
classical manifestation of QCD. It is this Skyrme approach
that we have taken to describe low energy proton antiproton
annihilation. We have shown that the
pion momentum, number and charge spectra emerge correctly
from this picture \cite{ACDLS} and that extending the Skyrme
treatment to include the classical omega field makes it
possible to also account for the omega mesons seen in
low energy annihilation \cite{bs&rda}.

In this paper we extend the large $N_C$
picture further to include the rho meson and study its
effects on annihilation. We concentrate on the observables
of nucleon antinucleon annihilation at rest and show that
with very few parameters we are able to account for the
major features of the pion momentum and number spectrum,
the branching ratios into the many annihilation channels
and the
ratio of direct pions to those from vector meson
decay \cite{physLet}.  Two
principal features of our treatment seem to be crucial to
giving this agreement.  First, classical QCD generates the
correct fraction of vector mesons during the annihilation
process.  We do not have to assume the vector mesons were present in
the baryons prior to annihilation.  Second, the coherent state,
with the constraint of isospin and four momentum imposed,
describes all the annihilation channels in terms of a single
quantum state, with each channel obtained from projections on
that state.  Thus the different channels have a common origin
and the branching ratios come out naturally from that state
with the constraints of phase space already imposed.

Early studies of annihilation in the Skyrme approach
\cite{SSLK,SWA} showed that when a Skyrmion and
antiSkyrmion annihilate the classical pion field streams
away from the annihilation region as fast as causality
will permit, leaving in a coherent burst of energy.
Thus far no one has used the
Skyrme approach to model the colliding of a proton and
antiproton and its subsequent radiation.  There are no
real conceptual obstacles to doing such a calculation
but it is technically challenging, and it has not been done.
It is, of course, still far easier than the corresponding
quantum calculation in lattice QCD.  Our models of annihilation
have tried to exploit the rapid appearance of the pion
pulse and have studied the development of the pion wave
after the annihilation region has been formed.  We begin
with a ``blob" of pionic matter with total energy of two
nucleons, zero baryon number and in a region of about
one Fermi radius.  For simplicity we take this region to be
spherically symmetric. We then use the large $N_C$
dynamics to propagate
the pion field outward into the radiation zone, where
the nonlinearities of the dynamics can be neglected,
and use the classical pion wave in the radiation zone to
construct a coherent state. The observed features of the pions
from annihilation are then studied in terms of that state.
To obtain agreement with experiment it is necessary to
impose both isospin and energy-momentum conservation on the
coherent state.  We have done all this both in the pure Skyrme
picture, which includes pions only \cite{ACDLS},
and in a generalized form
that includes omega mesons \cite{bs&rda}.

In this work we extend the theory to include rho mesons.
They are included in the large $N_C$ dynamics in a standard way \cite{BZ}.
We also focus on the phenomenology of annihilation.
We find agreement with the major trends of branching
ratios and spectra with essentially no free parameters.
Thus including the rho meson not only yields
considerable phenomenological
improvement but it also  widens the circle of QCD dynamics that
can be modeled in this classical/coherent state way.
In the next section we briefly outline our plan of attack.
Section 3 gives the massive Yang-Mills formulation
for including rho and omega fields in the Skyrme
or large $N_C$ dynamics.
Section 4 details the initial state assumptions we make,
how the asymptotic fields are obtained and their form.
Section 5
gives the quantum coherent state calculation based on those
asymptotic classical fields. Section 6 gives our results
and Section 7 presents some conclusions and prospectives
for future work.

\section{Plan of Attack}

In low energy
$\overline{p}p\rightarrow {\rm n}\pi$  about 33\% of
pions come out not directly from the annihilation but
from the decay of meson resonances \cite{SS}.
These are mostly the
vector meson resonances, $\rho$ and $\omega$.
To introduce the vector mesons in large $N_c$ QCD, we treat
them as massive Yang-Mills fields which gauge the $U_V(2)$
symmetry of the non-linear $\sigma$ model (contains $\pi$ only).
This non-linear $\sigma$ model is the starting
point of the Skyrme model or of any other classical or
large $N_C$ treatment of QCD.
The chiral anomaly gives the coupling between $\omega
\rho\pi$, responsible for the interactions
$\omega\rightarrow\pi^+\pi^-\pi^0$,
$\omega\rightarrow\rho\pi$ etc.  The $SU(2)$ gauge coupling
leads to $\rho\rightarrow 2\pi$. Vector dominance is in the
model and the  baryon is
stabilized by the vector meson terms,
eliminating the need for the Skyrme term.
Though no one has derived classical QCD from quantum QCD,
it is generally believed that the classical picture will
begin with the non-linear sigma term, include the gauged
vector mesons and continue in a series of terms with
higher and higher derivatives.  These higher terms become
important at high momentum or short distance, but since
annihilation at rest is, in some sense, a low energy
process, we believe keeping only the first few terms in that
expansion is a reasonable starting point.  We further
believe our results give some support to that view.

With our classical field theory, we study the dynamics
of annihilation at rest by starting with a ``blob" of  matter
of baryon number zero, of size about $1$ fm and
of energy twice the nucleon mass, made up purely of pions.
We use the dynamical equations to study the
space-time evolution of this blob including the
generation of classical vector meson fields.
At large $t$ and $r$, the fields decouple and we have free
field solutions. We quantize these free waves by
coherent state techniques.  We project the coherent
state onto states of good four-momentum and good isospin.

This quantum coherent state, generated from our dynamical, classical
field solutions, is the starting point of our phenomenology.
We obtain the particle spectra and branching ratios from the
coherent state simply by taking the appropriate matrix element
projection of that state.  This means that all the various annihilation
channels originate from that single state and are thus related.
This relationship coupled with the correct phase space imposed
by four-momentum conservation, leads naturally to a unified and
satisfactory account of the data.

\noindent
\section{Massive Yang-Mills Fields: the Classical Solution}

We introduce the $\rho$ and $\omega$ classical fields as $U(2)_V$-gauge
fields in the non-linear $\sigma$ model. The Lagrangian for the
non-linear $\sigma$ model is
\begin{equation}
{\cal L}_0=\frac{f_{\pi}^2}{4}Tr\left[\partial_\mu U
\partial^\mu U^{\dagger}\right]
\end{equation}
where $U$ is a unitary field in isospin space and $r$ space and where
$f_{\pi}$ is the pion decay constant.  This non-linear $\sigma$ term
is also the leading part of the Skyrme Lagrangian \cite{Skyrme}.
We introduce the $\rho$ and $\omega$ fields as massive gauge fields
through  the $U(2)_V$ covariant derivative
\begin{equation}
D_{\mu}U=\partial_{\mu}U-ig\left[\frac{\bfx{\tau}}{2}\cdot
\bfx{\rho}_\mu,\;U\right] \label{covariant_derivative}
\end{equation}
where $\bfx{\tau}$ is the Pauli matrices for isospin and greek
indices specify space-time components.
(Note that isovectors are boldface while arrows and ``hat" denote spatial
vectors.)
Since the $\omega$ field commutes with $U$, it does not appear
in (\ref{covariant_derivative}).
Its couplings are given entirely by the gauged Wess--Zumino anomaly term.
The
complete Lagrangian then is \cite{BZ,MK}
\begin{equation}
{\cal L}= {\cal L}_\sigma + {\cal L}_{m} + \Gamma_{WZ}
\end{equation}
with
\begin{eqnarray}
{\cal L}_\sigma&=&\frac{f_\pi^2}{4}Tr\left[D_{\mu}U
D^\mu U^{\dagger}\right]
+\frac{1}{2} m_{\pi}^2 f_{\pi}^2 Tr[U-1]
-\frac{1}{4}\left(\bfx{\rho}^2_{\mu\nu}+\omega_{\mu\nu}^2\right) ,
\\
{\cal L}_m&=&\frac{m^2}{2}\left(\bfx{\rho}_\mu^2+\omega_\mu^2\right) ,
\\
\Gamma_{WZ}&=& \Gamma^0_{WZ}[U] +\frac{3}{2}g \omega_\mu B^{\mu}
\nonumber \\
& &
+ \frac{3 g^2}{64\pi^2}\epsilon^{\mu\nu\alpha\beta}
\omega_{\mu\nu}Tr\left[ i \bfx{\tau}\cdot \bfx{\rho_\alpha}
\left( U^{\dagger}\partial_\beta U+\partial_\beta U U^{\dagger}\right)
\right.
\nonumber \\
& &
+ \left.\frac{g}{2}\bfx{\tau}\cdot\bfx{\rho_\alpha}U^{\dagger}
\bfx{\tau}\cdot\bfx{\rho_\beta}U\right]
\end{eqnarray}
where ${\cal L}_\sigma$ is the gauged nonlinear $\sigma$ model with a
term for the pion mass.
 The field strength tensors of the vector mesons
are given by
\begin{eqnarray}
\bfx{\rho}_{\mu\nu}&=&
\partial_{\mu} \bfx{\rho}_\nu-
\partial_{\nu} \bfx{\rho}_\mu + g \bfx{\rho}_\mu\times\bfx{\rho}_\nu ,
\nonumber \\
\omega_{\mu\nu}&=&
\partial_{\mu} \omega_\nu-
\partial_{\nu} \omega_\mu .
\end{eqnarray}
The mass term for the vector mesons is ${\cal L}_m$ and
we relate the ``charge" $g$ to the pion decay constant and the
vector meson mass (assumed to be the same for $\rho$ and $\omega$), $m=770$
MeV by the KSFR relation \cite{KSFR},
$ m= \sqrt{2} f_{\pi} g$.
The gauged Wess-Zumino term gives $\Gamma_{WZ}$.
The baryon current is given by
\begin{equation}
B^{\mu}=\frac{1}{24\pi^{2}}\epsilon^{\mu\nu\alpha\beta}
Tr\left[ (U^{\dagger}\partial_{\nu}U)(U^{\dagger}\partial_{\alpha}U)
(U^{\dagger}\partial_{\beta}U)\right]
\label{eq:baryon-current}
\end{equation}
and the Wess Zumino term for the pion field
$\Gamma_{WZ}^0(U)=0$ for $SU(2)$-valued
$U$ field.

We now wish to build in our initial condition of a spherically
symmetric blob of pionic matter.  That means that the pion field will
always be of the hedgehog type, and we may write
\begin{equation}
U=\exp\left[i\bfx{\tau}\cdot\hat{r}F(r,\;t)\right]
\end{equation}
where $F$ is essentially the pion field.
Furthermore we can decompose the vector meson fields into
a time and space components as follows
\begin{equation}
\omega^\mu=\left\{ \begin{array}{ll}
\omega_0(r,t) & \mu=0 \\
\hat{r}^i \omega_r(r,t) &\mu=i
\end{array}\right. \; \; \mbox{\rm{and}}\;\;
\rho^{\mu a}=\left\{ \begin{array}{ll}
\hat{r}^a H(r,\;t) & \mu=0 \\
\epsilon^{iak}\hat{r}^k \frac{G(r,\;t)}{gr}  & \mu=i
\end{array}\right. .
\end{equation}
Here $a$ is an isopin label and $\mu$  a space-time label.
The spatial unit vector is denoted by $\hat{r}$ and
the indices $a,i,j,k$ range over $1,2,3$.
We have explicitly removed
a $1/r$ from the spatial part of the rho field.  With our
initial conditions, the
$H$ field decouples and we will drop it in subsequent
equations.
In terms of these quantities, we can write
\begin{equation}
{\cal L}={\cal L}_{\pi}+{\cal L}_{\omega}+{\cal L}_{\rho}+{\cal
L}_{\pi\rho}+{\cal L}_{WZ}
\end{equation}
with
\begin{eqnarray}
{\cal L}_\pi&=& \frac{f_\pi^2}{2}
\left[
\left(\frac{\partial F}{\partial t}\right)^2
- \left(\frac{\partial F}{\partial r} \right)^2
-\frac{2}{r^2} \sin^2 F\right]
+f_\pi^2 m_\pi^2 (\cos F -1) ,
\\
{\cal L}_\omega&=& \frac{1}{2}\left(
\frac{\partial\omega_r}{\partial t} -\frac{\partial \omega_0}{\partial r}
\right)^2
+\frac{1}{2} m^2 ( \omega_0^2-\omega_r^2) ,
\\
{\cal L}_\rho&=&
\frac{1}{g^2 r^2}\left[
\left(\frac{\partial G}{\partial t}\right)^2
-
\left(
\frac{\partial G}{\partial r}\right)^2
-m^2 G^2
-\frac{1}{2 r^2} G^2 (G+2)^2\right] ,
\\
{\cal L}_{\pi\rho}&=&-f_\pi^2\frac{\sin^2 F}{r^2}G(G+2) ,
\\
{\cal L}_{WZ}&=&
-\frac{3g}{16\pi^2}\left[4
\left(\omega_0
\frac{\partial F}{\partial r}-
\omega_r
\frac{\partial F}{\partial t}\right)
\frac{\sin^2 F}{r^2} \right.
\nonumber \\
& & +
\left.
\left(
\frac{\partial \omega_0 }{\partial r}
-
\frac{\partial\omega_r}{\partial t}
\right) \frac{\sin 2 F}{r^2} G(G+2)\right] .
\end{eqnarray}
The nonlinearities of the theory and the nature of the
field couplings are clear in the various terms.
It should be stressed again that though we have put
$c$ the speed of light to one in these equations,
we have not had to do that for $\hbar$.  These are
purely classical equations, and there is no $\hbar$.

The equations of motion that follow from this Lagrangian are

\begin{eqnarray}
\left(\frac{\partial^2}{\partial t^2}-
\frac{\partial^2}{\partial r^2}-\frac{2}{r}\frac{\partial}
{\partial r}\right) F
 +  m_{\pi}^2 \sin F &=& - \frac{\sin 2F}{r^2} (G+1)^2
\nonumber \\ & &
- \frac{3g}{8\pi^2}\frac{1}{f_\pi^2 r^2}
\left(
\frac{\partial\omega_r}{\partial t}
-\frac{\partial\omega_0}{\partial r}
\right)
\nonumber \\ & &\left[2 \sin^2 F - \cos 2F \;G(G+2)\right] ,
\end{eqnarray}

\begin{eqnarray}
\left(\frac{\partial^2}{\partial t^2}-
\frac{\partial^2}{\partial r^2}-\frac{2}{r}\frac{\partial}{\partial r}
+\frac{2}{r^2}+m^2\right) \omega_r
&=&
\frac{3 g}{4\pi^2} \frac{\sin^2 F}{r^2}
\frac{\partial F}{\partial t}
-\frac{3g}{8\pi^2}\frac{1}{r^2}\left[ \cos 2F\; G(G+2)\frac{\partial F}
{\partial t}
\right.
\nonumber \\ & &
\left.
+ \sin 2F\; (G+1) \frac{\partial G}{\partial t}\right]
\end{eqnarray}
\begin{eqnarray}
\left(\frac{\partial^2}{\partial t^2}-
\frac{\partial^2}{\partial r^2}-\frac{2}{r}\frac{\partial}{\partial r}
+m^2\right) \omega_0&=&
\frac{3 g}{4\pi^2} \frac{\sin^2 F}{r^2}
\frac{\partial F}{\partial r}
-\frac{3g}{8\pi^2}\frac{1}{r^2}\left[ \cos 2F\; G(G+2)\frac{\partial F}
{\partial r}
\right.
\nonumber \\ & &
\left.
+ \sin 2F\; (G+1) \frac{\partial G}{\partial r}\right] ,
\end{eqnarray}

\begin{eqnarray}
\left(\frac{\partial^2}{\partial t^2}-
\frac{\partial^2}{\partial r^2}
+m^2\right) G + \frac{1}{r^2} G(G+1)(G+2)
&=&
 \frac{3g^3}{16\pi^2}\left(
\frac{\partial\omega_r}{\partial t}
-\frac{\partial\omega_0}{\partial r}
\right) \sin 2F\; (G+1)
\nonumber \\ &&
-g^2 f_\pi^2 \sin^2 F \; (G+1) .
\end{eqnarray}


The corresponding energy density is:
\begin{eqnarray}
{\cal H}&= & \frac{f_\pi^2}{2}
\left[
\left(\frac{\partial F}{\partial t}\right)^2
+ \left(\frac{\partial F}{\partial r} \right)^2
+\frac{2}{r^2} \sin^2 F\right]
+f_\pi^2 m_\pi^2 (1-\cos F )
\nonumber \\ &&
+\frac{1}{2}\left[
\left(\frac{\partial\omega_r}{\partial t} \right)^2
-\left(\frac{\partial \omega_0}{\partial r}\right)^2\right]
+\frac{1}{2} m^2 ( \omega_r^2 -\omega_0^2)
\nonumber \\ & &
+\frac{1}{g^2 r^2}\left[
\left(\frac{\partial G}{\partial t}\right)^2
+ \left(\frac{\partial G}{\partial r}\right)^2
+m^2 G^2
+\frac{1}{2 r^2} G^2 (G+2)^2\right]
\nonumber \\ & &
+f_\pi^2\frac{\sin^2 F}{r^2}G(G+2)
\nonumber \\ &&
+\frac{3g}{4\pi^2} \omega_0
\frac{\partial F}{\partial r}
\frac{\sin^2 F}{r^2}
+\frac{3g}{16\pi^2}
\frac{\partial \omega_0 }{\partial r}
\frac{\sin 2 F}{r^2} G(G+2) .
\label{eq_energy_density}
\end{eqnarray}

We use these equations to propagate our initial field
configuration into the radiation zone.  In the next
section we turn to a discussion of that configuration.

\noindent
\section{Initial Field Configurations and
Asymptotic Amplitudes}
As in \cite{bs&rda}, we model the initial pion field as
\begin{equation}
F(r,t=0)=h\frac{r}{r^{2}+a^{2}}\exp(-r/a)
\label{eq:initialF}
\end{equation}
with $a$ a range parameter and $h$  adjusted so that the total
energy equals to twice the nucleon mass.
In our previous studies we arbitrarily fixed  the
range by $a=1/m_{\pi}$.  Here, since we are interested in
a better treatment of the phenomenology, we take $a$ as a
parameter and fit it to obtain the correct asymptotic one
pion momentum distribution.  We find that $a=1.1$ fm leads
to an excellent account of the observed pion momentum distribution.
(see Section 6)
The initial field configurations are assumed to be
static, so that at $t=0$
\begin{equation}
\dot{F}=\dot{\omega}_{r}=\dot{\omega}_{0}=\dot{G}=0.
\end{equation}
We also take the initial vector meson fields to be zero. That is we take
\begin{equation}
\omega_{r}(r,t=0)=G(r,t=0)=0.
\end{equation}
Under these conditions we find
\begin{equation}
\omega_{0}(r,t=0)=\int_{0}^{\infty}dr^{'}G(r,r^{'})\left[-\beta r^{'2}
B^{0}(r^{'})\right]
\end{equation}
where
\begin{equation}
G(r,r^{'})=\frac{1}{2mrr^{'}}\left( e^{-m|r-r^{'}|}-e^{-m(r+r^{'})} \right)
\end{equation}
and $B^0(r)$ is the baryon density, calculated from (\ref{eq:baryon-current})
with the initial pion field (\ref{eq:initialF}).

Using the initial values we integrate the equations of motion
to determine the fields at later times.  As the fields propagate
outwards, they diminish in size, making the non-linear and coupling
terms in the Lagrangian less important.  Thus we can define a
radiation zone where the fields propagate as linear free massive fields.
In the radiation zone, we can calculate the pion,
$f(k)$, rho, $g(k)$, and omega, $h(k)$, momentum
distribution amplitudes from the expressions
\begin{equation}
\frac{d N_{\pi}}{d^{3}k}=\left| f(k) \right|^{2}=
\frac{1}{\pi k_{0}^{\pi} }
f_{\pi}^{2} \left| \int_{0}^{\infty} dr r^{2} j_{1}(kr)
(k_{0}^{\pi} + i\frac{\partial}{\partial t}) F(r,t) \right|^{2},
\label{eq:pionDist}
\end{equation}
\begin{equation}
\frac{d N_{\rho}}{d^{3}k}=\left| g(k) \right|^{2}=
\frac{2}{\pi k_{0}^{\rho} }
\left| \int_{0}^{\infty} dr r^{2} j_{1}(kr)
(k_{0}^{\rho}+i\frac{\partial}{\partial t})\frac{G(r,t)}{gr} \right|^{2},
\label{eq:rhoDist}
\end{equation}
and
\begin{equation}
\frac{d N_{\omega}}{d^{3}k}=\left| h(k) \right|^{2}=
\frac{1}{\pi k_{0}^{\omega} }
\left| \int_{0}^{\infty} dr r^{2} j_{1}(kr)
(k_{0}^{\omega}+i\frac{\partial}{\partial t})\omega_{r}(r,t) \right|^{2},
\label{eq:omegaDist}
\end{equation}
respectively.
Note that in extracting the fields we use that fact that
they all satisfy p-wave equations in the asymptotic region.

Let us examine the results of solving the dynamical equations.  In
Figure 1 we show the
dimensionless  pion field, $F(r,t)$, as it propagates out
to a time of 4 fm ($c=1$).  We see it begin as in (22) and diminish
as $1/r$.  It does not disperse very much because the pion mass is
small on the  scale of energies of the problem.  In Figure 2 we show
a similar plot of the rho field.  We plot the field,
$G(r,t)/gr$ reinstating the $1/gr$ factored out in (10).  This field
begins at $t=0$ as zero and builds  from the non-linear couplings before
beginning to diminish. The couplings and the large value of the
$\rho$ mass, make its behavior more complex than that of $F$.
Figure 3 shows the radial part of the
omega field, about which similar remarks apply.
Figure 4 shows the time component of the omega field multiplied
by $r$.  This makes the plotted quantity dimensionless.  We see
that it is much smaller than $F$, and also that the factor of
$r$ makes it not decrease for large $r$, since the fields themselves
all fall as $1/r$ for large $r$.

A more intuitive picture of the time development of the classical  fields
comes from studying the evolution of the energy density.  As we see from
(\ref{eq_energy_density}), that energy density has many parts.
Also for our choice of
initial condition all contributions to the energy density are spherically
symmetric.  Thus we do not plot the energy density as a function of
$r$ and $t$, but rather $4\pi r^2$ times that density.  The plotted
quantity need then only be integrated with $dr$ over all $r$ to obtain
the total energy.  In Figure 5 we  show the total radial energy density.  We
have checked that at each time slice the spatial integral of this density
gives twice the nucleon mass.  We see that the energy starts at $t=0$
rather compactly, propagates out and develops two ``humps." The faster of
these moves essentially along the light cone and is the energy in the
prompt pions. The slower hump is the energy in the massive vector mesons
and in the pions produced by the interaction of those vector mesons.
Figure 6 shows the radial energy density in the pions only. We see the
prompt peak on the light cone but the slow peak is smaller than that in
Figure 5 since much of that slow peak is not in pions.  Figures 7 and 8
show  the radial energy densities for the rho and omega fields.  These
are both slower, because of the mass, and smaller
(note  the scale) than the corresponding
pion energy.  Figures 9, 10, 11 and 12 show various terms in the
interaction energy.  These are small (again note the scales) and, more
importantly, decay quickly with increasing time or $r$.  It is this
diminished interaction energy at large $r$ that allows us to define
and asymptotic region where the fields propagate as free
massive fields.  It is these free fields that we use to construct
the coherent state.

\noindent
\section{Coherent State Calculation}
We use the asymptotic field strengths extracted in (\ref{eq:pionDist}),
(\ref{eq:rhoDist})
and (\ref{eq:omegaDist}) to construct quantum coherent states from the
classical $\pi$, $\rho$ and $\omega$ fields in the
radiation zone.  To obtain information on the pion
multiplicity from these states, we need to project
them into states of good four-momentum and isospin.  We do
this by known methods \cite{HornSilver,BSS}, that we
have elaborated previously \cite{ACDLS}.
First we construct
the field operators($F$, $G$ and $H$) which
create $\pi$, $\rho$ and $\omega$ at the space time position
$x$ and pointing in the isospin direction $\bfx{T}$. They are
\begin{equation}
F(x,\bfx{T})=\int d^3k \,f(\vec{k})\bfx{a}_{\vec{k}}^{\dagger}\cdot
\bfx{T}\,e^{-ik\cdot x},
\end{equation}
\begin{equation}
G(x,\bfx{T})=\int d^3k \,g(\vec{k})\bfx{b}_{\vec{k}}^{\dagger}\cdot
\bfx{T}\,e^{-ik\cdot x},
\end{equation}
and
\begin{equation}
H(x)=\int d^3k\, h(\vec{k})c_{\vec{k}}^{\dagger}\, e^{-ik\cdot x}.
\end{equation}
Note that $k\cdot x=k_0 t- \vec{k}\cdot\vec{x}$, with
$k_0=\sqrt{\vec{k}^2+M^2}$ ( $M$ is the corresponding meson mass).

{}From these we can form
the coherent state with $\pi$, $\rho$ and $\omega$ with
4-momentum and isospin projection as
\begin{eqnarray}
|I,I_z,K\rangle&=&\int \frac{d^4x}{(2\pi)^4}\frac{d\bfx{T}_1d\bfx{T}_2}{4\pi}
{\cal Y}^{\ast} _{II_z}(\bfx{T}_1,\bfx{T}_2)
\nonumber \\
& &\times e^{i K\cdot x} [e^{F(x,\bfx{T}_1)+G(x,\bfx{T}_2)+H(x)}-
F(x,\bfx{T}_1)-G(x,\bfx{T}_2)-H(x)-1]|0\rangle
\label{coherentstate}
\end{eqnarray}
and we have defined a set of coupled harmonic functions
of the $\pi$ and $\rho$ isospin directions
\begin{equation}
{\cal Y}_{II_z}(\bfx{T}_1,\bfx{T}_2)=
\sum_{I_1 M_1;I_2 M_2}\langle II_z|I_1 M_1; I_2 M_2\rangle
Y_{I_1M_1}(\bfx{T}_1)Y_{I_2M_2}(\bfx{T}_2).
\end{equation}

The states defined in (\ref{coherentstate})
are orthogonal
\begin{equation}
\langle I, I_z,K|I',I'_z, K'\rangle =\delta^4(K-K')\delta_{II'}
\delta_{I_zI_z'}{\cal I}(K)
\end{equation}
where the normalization factor is given by
\begin{eqnarray}
{\cal I}(K)=\int\frac{d^4x}{(2\pi)^4}
\frac{d\bfx{T}_1d\bfx{T}_2}{4\pi}
\frac{d\bfx{T}_1'd\bfx{T}_2'}{4\pi}
{\cal Y}_{II_z}(\bfx{T}_1,\bfx{T}_2)
{\cal Y}^{\ast}_{II_z}(\bfx{T}_1',\bfx{T}_2')
\nonumber \\
\times
e^{i K\cdot x} [e^{
\rho_f(x)\bfx{T}_1\cdot\bfx{T}_1'+
\rho_g(x)\bfx{T}_2\cdot\bfx{T}_2'+
\rho_h(x)
}
-\rho_f(x)\bfx{T}_1\cdot\bfx{T}_1'-
\rho_g(x)\bfx{T}_2\cdot\bfx{T}_2'-
\rho_h(x)
-1]|0\rangle
\end{eqnarray}
where
\begin{equation}
\rho_f(x)=\int d^3 p |f(\vec{p})|^2 e^{-ip\cdot x},
\end{equation}
\begin{equation}
\rho_g(x)=\int d^3 p |g(\vec{p})|^2 e^{-ip\cdot x}
\end{equation}
and
\begin{equation}
\rho_h(x)=\int d^3 p |h(\vec{p})|^2 e^{-ip\cdot x}.
\end{equation}

It is easy to see that the normalization integral is independent
of $I_z$.
We use the expansion method developed in \cite{ACDLS,bs&rda}
to calculate the normalization integral
\begin{equation}
{\cal I}(K)=\sum_{N_{\pi}+N_{\rho}+N_{\omega}\ge 2}
\frac{I(K, N_\pi, N_{\rho}, N_\omega)}{N_\pi! N_\rho! N_\omega!}
F(I, N_\pi, N_\rho)
\end{equation}
where
\begin{equation}
I(K, N_{\pi}, N_{\rho}, N_{\omega})=
\int \delta^{4}(K
-\sum_{i=1}^{N_{\pi}}p_{i}
-\sum_{j=1}^{N_{\rho}}q_{i}
-\sum_{k=1}^{N_{\omega}}r_{k})
\prod_{i=1}^{N_{\pi}} d^{3}p_{i} |f(\vec{p}_{i})|^{2}
\prod_{j=1}^{N_{\rho}} d^{3}q_{j} |g(\vec{q}_{j})|^{2}
\prod_{k=1}^{N_{\omega}}d^{3}r_{k} |h(\vec{r}_{k})|^{2} \label{eq:phase}
\end{equation}
and
\begin{equation}
F(I, N_\pi, N_\rho)=
\int
\frac{d\bfx{T}_1d\bfx{T}_2}{4\pi}
\frac{d\bfx{T}_1'd\bfx{T}_2'}{4\pi}
{\cal Y}_{II_z}(\bfx{T}_1,\bfx{T}_2)
{\cal Y}^{\ast}_{II_z}(\bfx{T}_1',\bfx{T}_2')
(\bfx{T}_1\cdot\bfx{T}_1')^{N_\pi}
(\bfx{T}_2\cdot\bfx{T}_2')^{N_\rho}.
\end{equation}
One can then show that
\begin{equation}
F(I, N_\pi, N_\rho)=
\sum_{I_1I_2}
{\cal F}(N_{\pi}, I_1)
{\cal F}(N_{\rho}, I_2)
\end{equation}
where the sum is over $I_1$ and $I_2$ which can add up to $I$.
\begin{eqnarray}
{\cal F}(N_{\pi},I)&=&\int \frac{d\bfx{T}d\bfx{T}^{'}}{4\pi}
Y^{*}_{II_z}(\bfx{T})Y_{II_z}(\bfx{T}^{'})
(\bfx{T}\cdot\bfx{T}^{'})^{N_{\pi}}
\nonumber \\
& =&\left\{   \begin{array}{ll}
0 & I >  N_{\pi} \mbox{ or } I-N_{\pi} \mbox{ is odd} \\
\frac{ N_{\pi}! }
{ (N_{\pi}-I)!! (I+N_{\pi}+1)!! } & I\le N_{\pi} \mbox{ and }
I-N_{\pi} \mbox{ is even}.
\end{array}   \right.
\end{eqnarray}

The mean numbers of $\pi$ and $\rho$ of isospin type $\mu$ in the state
(\ref{coherentstate}) are given by
\begin{eqnarray}
N_{\pi\mu}&=&\frac{1}{\cal I} \int \frac{d^4 x}{(2\pi)^4}
\int
\frac{d\bfx{T}_1d\bfx{T}_2}{4\pi}
\frac{d\bfx{T}_1'd\bfx{T}_2'}{4\pi}
{\cal Y}_{II_z}(\bfx{T}_1,\bfx{T}_2)
{\cal Y}^{\ast}_{II_z}(\bfx{T}_1',\bfx{T}_2')
T_{1\mu} T_{1\mu}^{'\ast}
\nonumber \\
&&
\times
 e^{iK\cdot x}\rho_f(x)
(e^{
\rho_f(x)\bfx{T}_1\cdot\bfx{T}_1'+
\rho_g(x)\bfx{T}_2\cdot\bfx{T}_2'+
\rho_h(x)
}-1)
\end{eqnarray}
and
\begin{eqnarray}
N_{\rho\mu}&=&\frac{1}{\cal I} \int \frac{d^4 x}{(2\pi)^4}
\int
\frac{d\bfx{T}_1d\bfx{T}_2}{4\pi}
\frac{d\bfx{T}_1'd\bfx{T}_2'}{4\pi}
{\cal Y}_{II_z}(\bfx{T}_1,\bfx{T}_2)
{\cal Y}^{\ast}_{II_z}(\bfx{T}_1',\bfx{T}_2')
T_{2\mu}T_{2\mu}^{'\ast}
\nonumber \\
&&
\times
e^{iK\cdot x}\rho_g(x)
(e^{
\rho_f\bfx{T}_1\cdot\bfx{T}_1'+
\rho_g(x)\bfx{T}_2\cdot\bfx{T}_2'+
\rho_h(x)
}-1) ,
\end{eqnarray}
respectively. The mean number of $\omega$ is given
by
\begin{eqnarray}
N_{\omega}&=&\frac{1}{\cal I} \int \frac{d^4 x}{(2\pi)^4}
\int
\frac{d\bfx{T}_1d\bfx{T}_2}{4\pi}
\frac{d\bfx{T}_1'd\bfx{T}_2'}{4\pi}
{\cal Y}_{II_z}(\bfx{T}_1,\bfx{T}_2)
{\cal Y}^{\ast}_{II_z}(\bfx{T}_1',\bfx{T}_2')
\nonumber \\ && \times
e^{iK\cdot x}\rho_h(x)
(e^{
\rho_f(x)\bfx{T}_1\cdot\bfx{T}_1'+
\rho_g(x)\bfx{T}_2\cdot\bfx{T}_2'+
\rho_h(x)
}-1).
\end{eqnarray}

Using expansion method, we can show that
\begin{equation}
N_{\pi\mu}=\frac{1}{\cal I}
\sum_{N_{\pi}+N_{\rho}+N_{\omega}\ge 1}
\frac{I(K, N_\pi+1, N_{\rho}, N_\omega)}{N_\pi! N_\rho! N_\omega!}
H(I, I_z, N_\pi,N_\rho,\mu)
\end{equation}
and
\begin{equation}
N_{\rho\mu}=\frac{1}{\cal I}
\sum_{N_{\pi}+N_{\rho}+N_{\omega}\ge 1}
\frac{I(K, N_\pi, N_{\rho}+1, N_\omega)}{N_\pi! N_\rho! N_\omega!}
H(I, I_z, N_\rho,N_\pi,\mu).
\end{equation}
The isospin factor is
\begin{eqnarray}
H(I,I_z,N_\rho,N_\pi,\mu)&=&\sum_{I_1I_1'I_2M_1}
\langle II_z|I_1M_1;I_2\;  I_z-M_1\rangle
\langle II_z|I_1'M_1;I_2\;  I_z-M_1\rangle
\nonumber \\
& &
\times
G(N_1, I_1, I_1', M_1,\mu){\cal F}(N_2,I_2)
\end{eqnarray}
with
\begin{eqnarray}
G(N, I_1, I_2, M_1,\mu)&=&\int\frac{d\hat T d\bfx{T}'}{4\pi}
Y^\ast_{I_1M_1}(\bfx{T})\bfx{T}_\mu^\ast \bfx{T}'_\mu (\bfx{T}\cdot
\bfx{T}')^N Y_{I_2M_1} (\bfx{T}')
\nonumber \\
&=&
\sum_{LM}\frac{2L+1}{\sqrt{(2I_1+1)(2I_2+1)}}
\langle I_10|10;L0\rangle
\langle I_20|10;L0\rangle
\nonumber \\
&&
\times
\langle I_1M_1|1\mu;LM\rangle
\langle I_2M_1|1\mu;LM\rangle .
\end{eqnarray}
Similarly for the $\omega$ we have
\begin{equation}
N_{\omega}=\frac{1}{\cal I}
\sum_{N_{\pi}+N_{\rho}+N_{\omega}\ge 1}
\frac{I(K, N_\pi, N_{\rho}, N_\omega+1)}{N_\pi! N_\rho! N_\omega!}
F(N_\pi,N_\rho,I,I_z).
\end{equation}
The probability of finding $N_\pi$ pions, $N_\rho$ rhos and $N_\omega$
omegas is given by
\begin{equation}
p(N_\pi,N_\rho,N_\omega)=\frac{1}{{\cal I}(K)}\frac{I(K,N_\pi, N_\rho,
N_\omega)}{N_\pi!N_\rho!N_\omega!}F(I, N_\pi,N_\rho)
\label{eq:prob}
\end{equation}
The probability of finding exactly $n$ pions, from
direct pions and from $\rho$ and $\omega$ decays, can be obtain
from the joint probability (\ref{eq:prob})
\begin{equation}
P_n=\sum_{n=N_\pi+2 N_\rho +3 N_\omega}p(N_\pi,N_\rho,N_\omega)
\label{eq:Pn}
\end{equation}
In the same spirit, we can also obtain the probability of having
certain of number of mesons
of each type and of given charge in the coherent state.
That probability is
\begin{eqnarray}
p( N_{\pi^+}, N_{\pi^-}, N_{\pi^0}, N_{\rho^+}, N_{\rho^-}, N_{\rho^0},
N_{\omega})&=& \frac{1}{\cal I}
\frac{
I(K,N_{\pi^+}+N_{\pi^-}+N_{\pi^0},
N_{\rho^+}+N_{\rho^-}+N_{\rho^0}, N_\omega)}
{
N_{\pi^+}!
N_{\pi^-}!
N_{\pi^0}!
N_{\rho^+}!
N_{\rho^-}!
N_{\rho^0}!
N_{\omega}!
}
\nonumber \\
& &
\times
F(I, I_z,
N_{\pi^+},
N_{\pi^-},
N_{\pi^0},
N_{\rho^+},
N_{\rho^-},
N_{\rho^0})
\end{eqnarray}
where
\begin{eqnarray}
F(I,I_z,
N_{\pi^+},
N_{\pi^-},
N_{\pi^0},
N_{\rho^+},
N_{\rho^-},
N_{\rho^0})&=&
\int
\frac{
d\hat{T_1}d\hat{T_2}
d\hat{T_1}'d\hat{T_2}'}
{(4\pi)^2}
{\cal Y}_{II_z}(\bfx{T}_1,\bfx{T}_2)
{\cal Y}^{\ast}_{II_z}(\bfx{T}_1',\bfx{T}_2')
\nonumber \\
& &
\times
(T_{1+} T_{1+}^{'\ast})^{N_{\pi^+}}
(T_{1-} T_{1-}^{'\ast})^{N_{\pi^-}}
(T_{10} T_{10}^{'\ast})^{N_{\pi^0}}
\nonumber \\
& &
\times
(T_{2+} T_{2+}^{'\ast})^{N_{\rho^+}}
(T_{2-} T_{2-}^{'\ast})^{N_{\rho^-}}
(T_{20} T_{20}^{'\ast})^{N_{\rho^0}} .
\end{eqnarray}
We now turn to the results from this formalism.

\noindent
\section{Results}
One of the advantages of our approach to annihilation is that
it contains very few parameters and that all annihilation
channels are determined at once by those few parameters.  Before
we can discuss results, we must fix those parameters.  The
pion mass we fix at its experimental value of $140$ Mev.
The vector mesons,
$\rho$ and $\omega$, are assigned the same mass of $m=770$ MeV.
We choose the parameters, $f_\pi$ and $g$, to satisfy the
KSFR relation \cite{KSFR} $m=\sqrt{ 2 } f_\pi g$.
We choose three values for $f_\pi$: 1) $93$ MeV (experimental);
2) $75$ MeV \cite{BZ}; 3) $62$ MeV \cite{AN}.
These last two values are popular choices in the
Skyrmion literature where they are determined by
getting the nucleon mass correct. In particular the
middle value, $f_{\pi} = 75$ MeV gives the
observed nucleon mass in our formalism with
gauged $\rho$ and $\omega$ fields \cite{BZ}. The corresponding
coupling constant $g$ is: 1) $5.85$; 2) $7.26$; 3) $8.78$.
The only remaining parameter is the size
of the initial spherical pion configuration. We fit that
to obtain the correct single pion momentum distribution,
giving $a=1.1$ fm. This is certainly a reasonable range.
All the remaining results to be discussed in this section
follow from those few, fixed parameters.

We begin with the single particle
inclusive pion momentum spectrum
from proton antiproton annihilation at rest.
This is shown in Figure 13 where the data is
compared with our calculation.  As discussed above,
one parameter in the initial pion configuration has
been adjusted to obtain the fit seen in Figure 13.  Nevertheless
we should stress that the entire machinery of our non-linear
classical dynamics and subsequent quantum coherent state
formalism intervenes between that initial configuration and
the asymptotic momentum distribution seen in Fig.13.  That
adjusting only one parameter in the initial configuration
is capable of giving such a good fit to the distribution
validates both the form of the initial configuration and
the subsequent dynamics. In Figures 14 and 15 we show the
corresponding momentum distribution for the rho and the omega.

Nucleon antinucleon annihilation can occur in two
isospin channels ($I=0,1$) with $I_z=0$ for $\bar{p} p$ and
$I_z=1 $ for $\bar{n} p$.  The mean number of mesons of each
charge type for the three isospin channels
 are shown in Tables 1, 2, and 3,
the different tables giving the
result for the different choices of $f_\pi$.
The charged pions numbers shown are
for the direct pions,
while the final pion average number includes both
direct pions and pions from vector meson decay.
We see that the average number of all pions and the width of
that distribution is rather insensitive to $f_\pi$, but the
details of the pion distribution and particularly the number
of vector mesons is strongly dependent on $f_\pi$.
In all cases the average pion number and the width agrees
well with the observed values. The results
in Table 3 use the same parameters as in our previous calculation
that did not include the $\rho$ meson \cite{bs&rda}. We see that including
the $\rho$ has slightly reduced the number of direct pions,
presumably to satisfy energy-momentum conservation.
In the remainder of this section we use only the middle
value of $f_{\pi}$ ($f_{\pi} = 75$ MeV)
since it is the one fit to the nucleon mass
in our classical field theory
\cite{BZ}.  Thus apart from the range parameter,
all our parameters are fixed by observed masses, pion, rho-omega
and nucleon.

In Figures 16 and 17 we show the pion number distribution
for each of the isospin channels.  These are
both for primary pions and for those from vector meson decay.
In each case we find the observed mean number of 5 with variance
of 1.  Also shown in those figures is a Gaussian distribution with
the same average and variance.  The Gaussian is nearly indistinguishable
from the our calculation.  Yet a Gaussian distribution is normally
taken as evidence for a statistical or fireball process, while our
process involves a rapid coherent pion wave. This should suggest a
cautionary note to attempts to draw lessons about annihilation
dynamics from the Gaussian pion number distribution.

In Table 4 we show the branching ratios for the many resolved channels
in proton antiproton annihilation at rest.  Note that some of the
channels involving more than two $\pi^0$ are not resolved. Note also
that the experimental signature of channels with vector mesons is
uncertain and there are   corresponding differences in the reported
ratios from different experimental groups. The branches involving
pions only in the second half of the table are for all
pions, primary and secondary. We can easily convert our branching ratios
for vector meson channels into pions since the vector mesons each have
a principal pion decay mode; $\rho^{\pm} \rightarrow \pi^{\pm} + \pi^0$,
$\rho^0 \rightarrow \pi^+ + \pi^- $ and $\omega \rightarrow \pi^+ + \pi^-
+ \pi^0$. In Table 4 we show our calculation
of the branching ratio for each of the isospin
channels separately, and then show the combined result based on an equal
mixture of $I=0$ and $I=1$.  Our results would hardly change if we took
the mix of 63\% $I=0$ and 37\% $I=1$ suggested by \cite{Milan}.

Also shown in the last row of Table 4 is the percentage of secondary pions.
That is the fraction of pions coming from resonance decays.  These are
nearly all from $\rho$ and $\omega$ decay.  The experimental fraction is
33\%, and we find 30\%, in quite reasonable agreement.  The dynamical
generation of vector mesons by our classical QCD is crucial to this result.

The major features of the
branching ratio data are correctly reproduced by our calculation
with essentially no free parameters.  The small channels come out small
and the large large.  For many of the large channels the agreement is
excellent.  Our biggest relative failing is for the $\pi^+ \pi^-$ channel
and for the $\pi^+ \pi^- \pi^0$ channel. These are both two body  channels
since the $\pi^+ \pi^- \pi^0$ channel is dominated by $\pi \rho$.
We  would expect the quantum corrections to be largest for the channels
with the fewest quanta, and that is what we are seeing.  Nevertheless
the overall agreement lends considerable strength to the underlying
treatment of classical QCD to generate the fields and a quantum coherent
state projected on four momentum and isospin to give the
various channels.  It is the single quantum coherent state that is
critical to explaining all the annihilation channels in a unified
context. It is also important that four momentum conservation
carry the constraints of phase space.  These constraints are
clearly seen in the broad-brush division into large and small channel.

The effect of the classical field dynamics is most clearly seen
in the generation of
the observed vector mesons, since we start with zero
vector meson fields.  The first part of Table 4 shows that we
are getting the vector meson channels qualitatively correctly and this
is further verified in the correct fraction of secondary pions.
Another way to see that our dynamics matters and that our agreement
is not all from phase space is to do a pions only calculation.  We
construct a coherent state of pions only with an inclusive single
particle momentum distribution fit to the empirically observed one
\cite{Dover}. We impose isospin and four momentum conservation and
once again calculate the branching ratios to the pion channels.
The results of this calculation are shown in Table 5.  Note
that for pions only there is an odd even isospin effect. That
is states of odd numbers of pions can only contribute to $I=1$ and
even numbers to $I=0$ \cite{ACDLS}.  We see in Table 5 that both
the mean number of pions and the details of the branching ratios
are in far less good agreement with experiment than our
previous calculation reported in Table 4 and including the full
dynamics and the vector mesons.  We conclude therefore, that just
fitting the single pion spectrum and using it to construct a
coherent state with phase space constraints imposed is not good
enough.  The underlying QCD dynamics also matters.

\noindent
\section{Summary and Conclusion}

We have shown that beginning with a classical field theory based
in large $N_C$ QCD we can give a good, nearly parameter free account
of the major features of nucleon antinucleon annihilation at rest.  Our
premise is that annihilation occurs as a rapid process with a sudden
burst of radiation and that that radiation can be described classically.
We begin with the non-linear sigma model, the starting point
of any large $N_C$ QCD, and use gauge invariance to add classical
rho and omega fields to the classical pion field.   As these  fields
propagate out from the annihilation region, they decouple and can be
well approximated by free fields.  We use these asymptotic fields to
construct a quantum coherent state and project that state onto states
of good four momentum and isospin.  We then calculate annihilation
observables from that coherent state. We find good agreement with
particle spectra and branching ratios.  That agreement arises from
our unified treatment of all the annihilation channels in terms
of a single coherent state, which also has four momentum and isospin
conservation imposed. It is interesting to observe that the number
spectrum from the coherent states strongly resembles the experimental
Gaussian distribution, while the conventional statistical approach
\cite{Blumel} (under the constraint of conservation of charge,
isospin, etc.) failed to produce the Gaussian distribution.

Our agreement with data is better than one might expect, given our
simplified starting assumptions, but it is certainly not perfect.
In particular the need for quantum corrections in the annihilation
branches involving few final quanta is clear. The entire question of
quantum corrections is an interesting and difficult one, but one that
should now be seriously address in view of the demonstrated success of
the first cut theory based on classical fields.  We are presently
studying quantum enhancements of the two pion channel to address the
charge spin asymmetry seen
in proton antiproton annihilation at higher energies \cite{Hasan,ALK}.
Other leading quantum corrections are also under
investigation.
We are also investigating annihilation in flight.  This would build in
the two center nature of the process and allow a description of
Bose Einstein correlations \cite{ACDLL}.
Further afield one might imagine a description involving classical
$SU(3)_f$ fields to describe the small annihilation branches into
K-mesons.

The success in annihilation of the classical QCD starting point followed
by quantum coherent states to describe the observed particles should
give hope for similar methods in other problems where there are
many pions but where the energies are low and nonperturbative QCD
dominates the physics.  Such approaches have already been developed for
the disoriented chiral condensate \cite{DCC} and have been
suggested for heavy ion reactions \cite{B&D}.
All these situations raise the same questions.  First is classical
QCD appropriate for describing the dynamics and second can one
estimate or calculate the leading quantum corrections?
The difficulty of alternate approaches, the modest success reported
here for annihilation, and the promised volume of high energy heavy
ion data should make the investigation of these questions topical
and fruitful.

\section*{Acknowlegements}
This work was supported in part by the United States National
Science Foundation.

\newpage

\begin{table}[p]

\begin{center}
\begin{tabular}{|c||ccc|ccc|c||cc|}   \hline
Channel   &
$n_{\pi^+}$   & $n_{\pi^-}$   & $n_{\pi^0}$   &
$n_{\rho^+}$  & $n_{\rho^-}$  & $n_{\rho^0}$  &
$n_\omega$ & $n$ & $\sigma$ \\  \hline\hline
$I=0$ $I_z=0$  & 1.50 & 1.50 & 1.50
			   & 0.09 & 0.09 & 0.09
			   & 0.13 & 5.44 & 0.81 \\ \hline
$I=1$ $I_z=0$  & 0.99 & 0.99 & 1.95
			   & 0.16 & 0.16 & 0.20
			   & 0.13 & 5.32 & 0.73 \\ \hline
$I=1$ $I_z=1$  & 1.85 & 1.09 & 0.99
			   & 0.29 & 0.06 & 0.16
			   & 0.13 & 5.32 & 0.73 \\ \hline
\end{tabular}
\end{center}

\caption{Numbers of direct $\pi^+$, $\pi^-$ and $\pi^0$;
$\rho^+$, $\rho^-$ and $\rho^0$; $\omega$ for $f_\pi=93$ MeV. $n$
is the total number of pions (direct and decay products) and
$\sigma$ is the standard deviation calculated from $P_n$
(\protect{\ref{eq:Pn}}).}
\end{table}

\begin{table}[p]

\begin{center}
\begin{tabular}{|c||ccc|ccc|c||cc|}   \hline
Channel   &
$n_{\pi^+}$   & $n_{\pi^-}$   & $n_{\pi^0}$   &
$n_{\rho^+}$  & $n_{\rho^-}$  & $n_{\rho^0}$  &
$n_\omega$ & $n$ & $\sigma$ \\  \hline\hline
$I=0$ $I_z=0$  & 1.26 & 1.26 & 1.26
			   & 0.06 & 0.06 & 0.06
			   & 0.41 & 5.33 & 0.86 \\ \hline
$I=1$ $I_z=0$  & 0.80 & 0.80 & 2.09
			   & 0.10 & 0.10 & 0.14
			   & 0.33 & 5.39 & 0.74 \\ \hline
$I=1$ $I_z=1$  & 1.87 & 1.03 & 0.80
			   & 0.20 & 0.04 & 0.10
			   & 0.33 & 5.39 & 0.74 \\ \hline
\end{tabular}
\end{center}

\caption{Numbers of direct $\pi^+$, $\pi^-$ and $\pi^0$;
$\rho^+$, $\rho^-$ and $\rho^0$; $\omega$ for $f_\pi=75$ MeV.}
\end{table}

\begin{table}[p]

\begin{center}
\begin{tabular}{|c||ccc|ccc|c||cc|}   \hline
Channel   &
$n_{\pi^+}$   & $n_{\pi^-}$   & $n_{\pi^0}$   &
$n_{\rho^+}$  & $n_{\rho^-}$  & $n_{\rho^0}$  &
$n_\omega$ & $n$ & $\sigma$ \\  \hline\hline
$I=0$ $I_z=0$  & 0.93 & 0.93 & 0.93
			   & 0.02 & 0.02 & 0.02
			   & 0.75 & 5.18 & 0.82 \\ \hline
$I=1$ $I_z=0$  & 0.61 & 0.61 & 2.24
			   & 0.05 & 0.05 & 0.07
			   & 0.55 & 5.42 & 0.83 \\ \hline
$I=1$ $I_z=1$  & 1.88 & 0.96 & 0.61
			   & 0.10 & 0.02 & 0.05
			   & 0.55 & 5.42 & 0.83 \\ \hline
\end{tabular}
\end{center}

\caption{Numbers of direct $\pi^+$, $\pi^-$ and $\pi^0$;
$\rho^+$, $\rho^-$ and $\rho^0$; $\omega$ for $f_\pi=62$ MeV.}
\end{table}
\begin{table}[p]

\begin{center}
\begin{tabular}{|c|ccc|cc|}   \hline
     &
\multicolumn{3}{c|}{Theory} & \multicolumn{2}{c|} {Experiment}
\\
Channel & $I=0$ & $I=1$ & Combined & CERN & BNL \\ \hline
$\pi^0\rho^0$ & 0.013 & 0 & 0.006 & $3.9\pm0.4$ &  \\ \hline
$\pi^{\pm}\rho^{\mp}$ & 0.013 & 0 & 0.006 & $1.9\pm0.3$ &  \\ \hline
$\rho^0\pi^+\pi^-$ & 0 & 3.0 & 1.5 & $1.5\pm0.3$ & $5.8\pm0.3$
\\  \hline
$\rho^0\rho^0$ & 0.146 & 0 & 0.073 & $0.12\pm0.12$ & $0.4\pm0.3$
\\  \hline
$\pi^0\omega$ & 0.0 & 1.2 & 0.6 & $2.3\pm0.2$ & $0.7\pm0.2$  \\ \hline
$\omega\omega$ & 3.12 & 0 & 1.56 & $1.4\pm0.6$
\protect{\cite{Amsler}}&   \\ \hline
$\rho^0\pi^+\pi^-\pi^0$ & 2.0 & 0 & 1.0&  & $7.3\pm1.7$  \\ \hline
$\rho^{\pm}\pi^{\mp}\pi^+\pi^-$ & 7.9 & 10.3 & 9.1 &  &
$6.4\pm1.8$ \\ \hline
$\omega \pi^+\pi^-$ & 1.7 & 0 & 0.8 & $6.6\pm0.35$ &
$3.8\pm0.4$  \\ \hline
$\omega 2\pi^+2\pi^-$ & 4.3 & 0 & 2.1 & $1.3\pm0.3$ & \\  \hline
\hline
$\pi^+\pi^-$        & 0.02 & 0.0  & 0.01 &
$0.37\pm0.3$ &$0.32\pm0.04$ \\ \hline
$\pi^+\pi^-\pi^0$   & 0.04 & 0.6  & 0.32 &
$6.9\pm0.35$ & $7.3\pm0.9  $ \\ \hline
$2\pi^+2\pi^-$        & 9.1 & 3.0 & 6.1    &
$6.9\pm0.6$ & $5.8\pm0.3  $ \\ \hline
$2\pi^+2\pi^-\pi^0$ & 26.8 & 19.8 & 23.3 &
$19.6\pm0.7$ & $18.7\pm 0.9$ \\ \hline
$3\pi^+3\pi^-$       & 13.8 & 3.56 & 8.7 &
$2.1\pm0.2$ & $1.9\pm0.2$ \\ \hline
$3\pi^+3\pi^-\pi^0$ &  4.38 & 0.61 & 2.5&
$1.9\pm0.2$ & $1.6\pm0.2$ \\ \hline
${\rm n}\pi^0$, ${\rm n}>1$ & 7.7 & 15.7 & 11.7 &
$4.1\pm0.4$ & $3.3\pm0.2$ \\ \hline
$\pi^+\pi^-{\rm n}\pi^0$, ${\rm n}>1$ & 25.1 & 39.8 & 32.5 &
$ 35.8\pm0.8$ & $34.5\pm1.2$ \\ \hline
$2\pi^+2\pi^-{\rm n}\pi^0$, ${\rm n}>1$ & 12.8 & 17.4 & 15.2 &
$ 20.8\pm0.7$ & $21.3\pm 1.1$ \\ \hline
$3\pi^+3\pi^-{\rm n}\pi^0$, ${\rm n}>1$ & 0.03 & 0.014 & 0.022 &
$0.3\pm0.1$ & $0.3\pm0.1$\\ \hline\hline
\% of secondary $\pi$s & 29.2 & 31.3 & 30.3
& \multicolumn{2}{c|}{33} \\ \hline
\end{tabular}
\end{center}

\caption{Branching ratios, in percent, for proton antiproton annihilation
at rest.  Our calculations are compared with experiments from
\protect{\cite{SS}}.
We show each total isospin channel calculated separately. The ``combined"
column corresponds to equal mixture of $I=0$ and $I=1$.  In the last
row we list the percentage of pions from the decay of rho and omega mesons.}
\end{table}
\begin{table}[p]

\begin{center}
\begin{tabular}{|c|ccc|cc|}   \hline
     &
\multicolumn{3}{c|}{Pion-only Calculation} & \multicolumn{2}{c|} {Experiment}
\\
Channel & $I=0$ & $I=1$ & Combined & CERN & BNL \\
\hline
$\pi^+\pi^-$        & 0.0004 & 0.0  & 0.0002 &
$0.37\pm0.3$ &$0.32\pm0.04$ \\ \hline
$\pi^+\pi^-\pi^0$   & 0 & 0.05  & 0.03 &
$6.9\pm0.35$ & $7.3\pm0.9  $ \\ \hline
$2\pi^+2\pi^-$        & 1.93 & 0 & 0.97   &
$6.9\pm0.6$ & $5.8\pm0.3  $ \\ \hline
$2\pi^+2\pi^-\pi^0$ & 0 & 7.0 & 3.5 &
$19.6\pm0.7$ & $18.7\pm 0.9$ \\ \hline
$3\pi^+3\pi^-$       & 37.2 & 0 & 18.6 &
$2.1\pm0.2$ & $1.9\pm0.2$ \\ \hline
$3\pi^+3\pi^-\pi^0$ &  0 & 10.5 & 5.2&
$1.9\pm0.2$ & $1.6\pm0.2$ \\ \hline
${\rm n}\pi^0$, ${\rm n}>1$ & 14 & 36 & 25 &
$4.1\pm0.4$ & $3.3\pm0.2$ \\ \hline
$\pi^+\pi^-{\rm n}\pi^0$, ${\rm n}>1$ & 16.8 & 30 & 23 &
$ 35.8\pm0.8$ & $34.5\pm1.2$ \\ \hline
$2\pi^+2\pi^-{\rm n}\pi^0$, ${\rm n}>1$ & 20.9 & 16 & 18.5 &
$ 20.8\pm0.7$ & $21.3\pm 1.1$ \\ \hline
$3\pi^+3\pi^-{\rm n}\pi^0$, ${\rm n}>1$ & 3 & 1.2 & 2.1 &
$0.3\pm0.1$ & $0.3\pm0.1$\\ \hline\hline
$\overline{n}$, $\sigma$ & 6.22, 0.83 & 6.4, 0.95  &
& \multicolumn{2}{c|}{5, 1} \\ \hline
\end{tabular}
\end{center}

\caption{Branching ratios, in percent, for proton antiproton annihilation
at rest, calculated from pions only, with experimental pion momentum
spectrum. Data are compared with experiments from
\protect{\cite{SS}}.
We show each total isospin channel calculated separately. The ``combined"
column corresponds to equal mixture of $I=0$ and $I=1$.  In the last
row we list the average and standard variation in the number of pions .}
\end{table}

\clearpage

\newpage
\setlength{\textwidth}{465pt}
\begin{figure}[p]
\centerline{\hbox{
}}

\caption{Pion field configuration F as a function of r and t.}
\end{figure}

\begin{figure}[p]
\centerline{\hbox{
}}

\caption{$\rho=G/gr$ (in units of vector meson mass $m$)
field as a function of r and t.}
\end{figure}

\newpage

\begin{figure}[p]
\centerline{\hbox{
}}

\caption{$\omega_r$ (in units of vector meson mass $m$)
field as a function of r and t.}
\end{figure}

\begin{figure}[p]
\centerline{\hbox{
}}

\caption{$u=r\omega_0$
field as a function of r and t.}
\end{figure}

\newpage

\begin{figure}[p]
\centerline{\hbox{
}}

\caption{Total energy density multiplied by $4\pi r^{2}$
(in units of MeV/fm) as a function of r and t.}
\end{figure}

\begin{figure}[p]
\centerline{\hbox{
}}

\caption{ Pion energy density multiplied by $4\pi r^{2}$
(in units of MeV/fm) as a function of r and t.}
\end{figure}

\newpage

\begin{figure}[p]
\centerline{\hbox{
}}

\caption{Energy density associated
with $\rho$ multiplied by $4\pi r^{2}$
(in units of MeV/fm) as a function of r and t.}
\end{figure}

\begin{figure}[p]
\centerline{\hbox{
}}

\caption{Energy density associated
with $\omega_{r}$ multiplied by $4\pi r^{2}$
(in units of MeV/fm) as a function of r and t.}
\end{figure}

\newpage

\begin{figure}[p]
\centerline{\hbox{
}}

\caption{Energy density associated
with $\rho$ self-interaction multiplied by $4\pi r^{2}$
(in units of MeV/fm) as a function of r and t.}
\end{figure}

\begin{figure}[p]
\centerline{\hbox{
}}

\caption{Energy density associated with $\pi\rho$ interaction
multiplied by $4\pi r^{2}$
(in units of MeV/fm) as a function of r and t. }
\end{figure}

\newpage

\begin{figure}[p]
\centerline{\hbox{
}}

\caption{Energy density associated with $\pi\omega$ interaction
multiplied by $4\pi r^{2}$
(in units of MeV/fm) as a function of r and t. }
\end{figure}

\begin{figure}[p]
\centerline{\hbox{
}}

\caption{Energy density associated with $\pi\rho\omega$ interaction
multiplied by $4\pi r^{2}$
(in units of MeV/fm) as a function of r and t. }
\end{figure}

\newpage

\begin{figure}[p]
\centerline{\hbox{
}}

\caption{Pion momentum distribution.
The horizontal axis
is the pion momentum.
The vertical axis is the pion momentum
distribution function, $4\pi k^2 |f(k)|^2$, see
(\protect{\ref{eq:pionDist}}).
The area under the curve gives the total
direct pion number $N_{\pi}$. The data is from
\protect{\cite{SS}}.
}
\end{figure}

\begin{figure}[p]
\centerline{\hbox{
}}

\caption{Rho momentum distribution.
The horizontal axis
is the rho momentum.
The vertical axis is the rho momentum
distribution function, $4\pi k^2 |g(k)|^2$, see
(\protect{\ref{eq:rhoDist}}).
The area under the curve gives the total
omega number $N_{\rho}$.}
\end{figure}
\newpage

\begin{figure}[p]
\centerline{\hbox{
}}

\caption{Omega momentum distribution.
The horizontal axis
is the omega momentum.
The vertical axis is the rho momentum
distribution function, $4\pi k^2 |h(k)|^2$, see
(\protect{\ref{eq:omegaDist}}).
The area under the curve gives the total
omega number $N_{\omega}$.}
\end{figure}

\begin{figure}[p]
\centerline{\hbox{
}}

\caption{Pion number distribution in $I=0$ channel.
Solid squares are given by the distribution $P_{n}$
(\protect{\ref{eq:prob}}).
Circles are the gaussian distribution with the same mean and variance.}
\end{figure}

\newpage

\begin{figure}[p]
\centerline{\hbox{
}}

\caption{Pion number distribution in $I=1$ channel.
Solid squares are given by the distribution $P_{n}$
(\protect{\ref{eq:prob}}).
Circles are the gaussian distribution with the same mean and variance.}
\end{figure}

\begin{thebibliography}{99}
\bibitem{'tHooft}
G. 't Hooft, Nucl.~Phys.~{\bf B 72}, 461 (1974); {\bf B 75}, 461 (1974).
\bibitem{Witten}
E.~Witten, Nucl.~Phys.~{\bf B 160}, 57 (1979).
\bibitem{DCC}
R.D.~Amado and I.I.~Kogan, Phys.~Rev.~{\bf D 51}, 190, 1995,
and references therein.
\bibitem{B&D}
J.-P. Blaizot and D. Diakonov, Phys.~Lett.~{\bf B 315}, 226 (1993).
\bibitem{ACDLS}
 R.D. Amado, F. Cannata, J-P. Dedonder, M.P. Locher, and B. Shao,
 Phys. Rev. {\bf C 50}, 640 (1994).
\bibitem{bs&rda}
B.~Shao and R.D.~Amado, Phys.~Rev.~{\bf C 50}, 1787 (1994).
\bibitem{Skyrme}
T.H.R.~Skyrme, Proc.~R.~Soc.~London {\bf 262}, 237 (1961);
Nucl.~Phys.~{\bf 31}, 556 (1962).
\bibitem{physLet}
A brief report of this work appeared in hep-ph/9504362, and was
submitted to Phys.~Lett.~{\bf B}.
\bibitem{SSLK}
H.M.~Sommermann, R.~Seki, S.~Larson and S.E.~Koonin,
Phys.~Rev.~{\bf D 45}, 4303 (1992).
\bibitem{SWA}
B.~Shao, N.R.~Walet and R.D.~Amado, Phys.~Lett.~{\bf B 303}, 1 (1993).
\bibitem{BZ}
I.~Zahed and G.E.~Brown, Phys.~Rep.~{\bf 142}, 1 (1986).
\bibitem{SS}
J.~Sedl\'ak and V.~\^Sim\'ak, Sov.~J.~Part.~Nucl.~{\bf 19} (3), 191 (1988).
\bibitem{MK}
U.G.~Mei{\ss}ner and N.~Kaiser, Z.~Phys.~{\bf A 325}, 267 (1986).
\bibitem{KSFR}
K.~Kawarabayashi and M.~Suzuki, Phys.~Rev.~Lett.~{\bf 16}, 255 (1966);
Riazzuddin and Fayyazuddin, Phys.~Rev.~{\bf 147}, 1071 (1966).
\bibitem{HornSilver}
D.~Horn and R.~Silver, Ann.~Phys.~{\bf 66}, 509 (1971).
\bibitem{BSS}
J.C.~Botke, D.J.~Scalapino and R.L.~Sugar,
Phys.~Rev.~{\bf D 9}, 813 (1974).
\bibitem{AN}
G.S.~Adkins and C.R.~Nappi, Phys.~Lett.~{\bf B 137}, 251 (1985).
\bibitem{Milan}
M.P.~Locher and B.S.~Zou, Z.~Phys.~{\bf A 341}, 191 (1992).
\bibitem{Dover}
C.B.~Dover, T.~Gutsche, M.~Maruyama and A.~Faessler,
Prog.~Part.~Nucl.~Phys.~{\bf 29}, 87 (1992).
\bibitem{Blumel}
W.~Bl\"umel and U.~Heinz, hep-ph-9409343, TPR-94-30.
\bibitem{Hasan}
A.~Hasan et al., Nucl.~Phys.~{\bf B 378}, 3 (1992).
\bibitem{ALK} R.D.~Amado, Yang Lu and I.I.~Kogan, in preparation.
\bibitem{ACDLL}
R.~D.~Amado, F.~Cannata, J.-P.~Dedonder, M.~P.~Locher and Yang Lu,
Phys.~Lett.~{\bf B 339}, 201 (1994); Phys.~Rev.~{\bf C 51}, 1587 (1995).
\bibitem{Amsler}
C.~Amsler and F.~Myhrer, Annu.~Rev.~Nucl.~Part.~Sci.~{\bf 41}, 219 (1991).
\end{thebibliography}
\end{document}